\documentclass[twocolumn,prl,superscriptaddress,longbibliography,floatfix]{revtex4-1}

\usepackage{graphicx}
\usepackage{amsmath}
\usepackage{amssymb}
\usepackage{bm}
\usepackage{times}

\usepackage[pdftex]{hyperref}
\hypersetup{colorlinks=true,linkcolor=blue,citecolor=blue,urlcolor=blue}

\begin{document}

\title{Why replica symmetry breaking does not occur below six dimensions in Ising spin glasses}

\author{M.~A.~Moore}
\affiliation{School  of
Physics and Astronomy, University of Manchester, Manchester M13 9PL, UK}

\date{\today}
\begin{abstract}
The leading term for the average height of the  barriers which could separate pure states in Ising spin glasses is calculated  using instanton methods.  It is finite in dimensions $d < 6$. Replica symmetry breaking requires that the barriers between pure states are infinite in the  thermodynamic limit, as  finite barriers allow  thermal mixing of pure states. As a consequence, a replica symmetry broken phase cannot survive when $d < 6$.  However, for $d > 6$ no similar instanton solution exists.
\end{abstract}

\maketitle

 The nature of the ordered phase of spin glasses has been controversial for decades. The standard calculational methods such as the renormalization group and mean-field theory are in disagreement with each other: The picture which derives from mean-field theory, which is valid for infinite dimensional systems, is that of broken replica symmetry (RSB) \cite{parisi:79,parisi:83,rammal:86,mezard:87,parisi:08}. This is contradicted by the results of real-space renormalization group (RG) and scaling calculations \cite{mcmillan:84a,bray:86,fisher:88}, which favor in low dimensions an ordered phase with replica symmetry \cite{moore:98,wang:17,angelini:15,angelini:17,angelini:17a}. Recent calculations using the strong disorder renormalization group suggest that the spin glass phase is replica symmetric for $d \le 6$ \cite{wang:17,wang:18b}.  Real-space RG calculations are ad  hoc, and it is hard to convince supporters of RSB with them. What is needed 
 is a  calculation within the replica framework  which shows why  RSB
will  go away below six  dimensions. To date, there  have been  been 
controversial hints \cite{Moore:11,moore:18} within the replica framework that six
might  be  the  special  dimension  below  which  RSB  might  not  exist, but see \cite{parisi:12,tamas:17,yaida:17}.

In the RSB state there are many pure states present whose free energies differ by $O(1)$. These states are separated by high barriers. Numerical studies \cite{Billoire:10,billoire:01} of the Sherrington-Kirkpatrick (SK) model \cite{sherrington:75}   and other arguments \cite{Rodgers:89,aspelmeier:06} suggest that these barriers depend on the number of spins $N$ in that model as $N^{1/3}$. It is vital to the whole RSB picture that the barriers between the pure states become infinite in the thermodynamic limit i.e. as $ N \to \infty$. If they are finite, thermal fluctuations will mix the pure states together and the RSB picture cannot then apply. The basic argument of this paper is that the barriers between the pure states of the RSB state would be finite when $d < 6$, thereby causing the whole RSB picture to fall apart.

We start from 
the Edwards-Anderson model \cite{edwards:75} defined on a
$d$-dimensional cubic lattice  with the  Hamiltonian
\begin{equation} 
\mathcal{H} = - \sum_{\langle ij \rangle} J_{ij} S_i S_j, 
\label{eq:ham} 
\end{equation} 
where the summation is over only nearest-neighbor bonds and the random
couplings $J_{ij}$ are  chosen from the standard Gaussian distribution
of unit variance and zero mean.  The Ising spins take the values $S_i
\in \{\pm 1\}$ with $i = 1,2, \ldots, N$.  From the expression for the partition function associated with Eq.~(\ref{eq:ham}) one can derive \cite{green:83,harris:76,pytte:79,bray:79} the  replicated  and bond-averaged functional in the fields $Q_{\alpha \beta}$  which is believed captures the essence of spin glass behavior in $d$ dimensions:
\begin{eqnarray}
&F[\{Q_{\alpha\beta}\}]/k_B T  =   \int   d^dr\,   \left[-\frac{1}{2} \tau
\sum_{\alpha < \beta}Q_{\alpha\beta}^2
+\nonumber \right. \\
& \left. \hspace{-0.5cm} \frac{1}{2}\sum_{\alpha < \beta}(\nabla
Q_{\alpha\beta})^2 
  - w
\sum_{\alpha <\beta < \gamma}Q_{\alpha\beta}
Q_{\beta\gamma}Q_{\gamma\alpha}  + O(Q^4)\right]
\label{FQ}
\end{eqnarray}
As usual for replicated systems the indices  $\alpha$, $\beta$ and $\gamma$ run over integer values $1, 2, \ldots, n$, and $n$ is set zero at the end of the calculation. In this limit, the averaged  free energy of the original system is
$F/n$, where $F$ is the replicated free energy, which has therefore to be proportional to $n$ as $n \to 0$. In the SK model one needs to include a quartic term $y Q_{\alpha \beta}^4$ in order to produce RSB. For the EA model we shall dispense with this term since the fluctuations around the mean-field solution cause RSB  \cite{bray:79}.  The coefficient $\tau$ vanishes at the mean-field transition temperature and is of the form $\sim (1-T/T_c)$. The gradient term weights the cost of having a spatially varying order parameter $Q_{\alpha \beta}$. We shall begin by outlining the old argument why replica symmetry apparently needs to be broken \cite{bray:79}.

Mean-field theory seeks stationary points of the free energy functional of Eq. (\ref{FQ}). We shall first examine the replica symmetric solution $Q_{\alpha \beta}(\mathbf{r})=Q$, which is independent of the replica indices $\alpha$ and $\beta$ and  does not depend upon the spatial position $\mathbf{r}$. This gives 
\begin{equation}
F(Q)/N k_B T =\frac{1}{2}n (n-1)[-\frac{1}{2} \tau Q^2 -\frac{1}{3}(n-2) w Q^3].
\label{fQmin}
\end{equation}
 $F(Q)$ is stationary when
\begin{equation}
-\tau Q-(n-2)w Q^2=0.
\label{Qstat}
\end{equation}
 There are two solutions: $Q=0$ and $Q=-\tau/(n-2)w$. For $\tau <0$, which corresponds to $T> T_c$, the trivial solution $Q=0$ is appropriate and describes the paramagnetic phase.  The appropriate solution for $\tau>0$, i.e. $T < T_c$ is the solution $Q=\tau/(2-n) w$. However, it appears to be unstable against fluctuations which break replica symmetry.

One can see this by writing 
\begin{equation}
q_{\alpha\beta}=Q+R_{\alpha\beta}
\label{Fl}
\end{equation}
and substituting  into Eq. (\ref{FQ}) (without the  quartic terms). Then
up to constants the free energy functional becomes
\begin{align}
&F[\{R_{\alpha\beta}\}]/k_BT         =   \int   d^dr\,   \left[   -  \frac         12 \tau
\sum_{\alpha<\beta}R^2_{\alpha\beta}+             \frac           12
\sum_{\alpha<\beta}(\nabla  R_{\alpha\beta})^2   \nonumber \right.  \\  &\left.  -wQ
\sum_{\alpha   <   \beta  <\gamma}   (R_{\alpha\beta}R_{\alpha\gamma}+
R_{\alpha\beta}R_{\beta\gamma}+R_{\alpha\gamma}R_{\beta\gamma}) 
\nonumber \right. \\      &\left.-w\sum_{\alpha       <        \beta       <\gamma}
R_{\alpha\beta}R_{\alpha\gamma}R_{\beta\gamma} \right] .
\label{q2}
\end{align} 
The quadratic terms are not  diagonal in the replica indices. To deal with this it is useful to first introduce
the  following   propagators  in  terms  of   the  Fourier  components
$R_{\alpha\beta}({\bf q})$
\begin{align}
&G_1(q)=\langle      R_{\alpha\beta}({\bf      q})R_{\alpha\beta}({\bf
  -q})\rangle,    \nonumber\\   &G_2(q)=\langle   R_{\alpha\beta}({\bf
  q})R_{\alpha\gamma}({\bf -q})\rangle,\hspace{0.5cm} \beta \ne \gamma
\nonumber\\            &G_3(q)=\langle            R_{\alpha\beta}({\bf
  q})R_{\gamma\delta}({\bf  -q})\rangle,  \hspace{0.5cm}  \alpha,\beta
\ne \gamma, \delta.
\label{defG}
\end{align}
Then,  following  Ref. \cite{bray:79}  the  quadratic  form is  readily
diagonalized in terms of three linear combinations of $G_1$, $G_2$ and
$G_3$:
\begin{align}
&G_B   \equiv   G_1+2(n-2)G_2+\frac  12(n-2)(n-3)G_3=(q^2+\tau)^{-1}
\nonumber\\     &G_A    \equiv    G_1+(n-4)G_2-(n-3)G_3=(q^2+2wQ)^{-1}
\nonumber\\ &G_R \equiv G_1-2G_2+G_3=(q^2+nwQ)^{-1}.
\label{rep}
\end{align}
All  three of these  propagators are  of the  form $(q^2+m^2_s)^{-1}$,
with the mass of the breather mode given by $m_L^2=\tau$, that of the
\lq anomalous'  mode by  $m_A^2=2wQ$ and that of the replicon  mode by
$m_R^2=nwQ$. In  the limit of $n  \rightarrow 0$ the  breather and the
anomalous masses  become equal while  the replicon mass goes  to zero.
Stabilty  of course  requires that  all the  $m^2_s$  be non-negative.
Thus  at Gaussian order  the replica  symmetric solution  has marginal
stability.  (If we  had retained the quartic terms  in the Hamiltonian
density  the replicon  mode  would have  become  unstable at  Gaussian
order).   To see  the apparent  instability of  the  replica symmetric
state in the absence of the quartic term  it is  necessary to  go  to one  loop order  and calculate  the
self-energies of  the propagators.  The replicon self-energy
$\Sigma_R(q)$ is defined via
\begin{equation}
G_R=(q^2+nwQ-\Sigma_R(q))^{-1}.
\label{SED}
\end{equation}

To one-loop order the  calculation of $\Sigma_R(q)$ is straightforward
\cite{bray:79,TDP};
  $\Sigma_R(0)$  is given by
\cite{bray:79}
\begin{equation}
\Sigma_R(0)          \approx          \frac{4w^2 \tau^2}{N}\sum_{{\bf
q}}\frac{1}{q^4(q^2+\tau)^2}.
\label{SEA}
\end{equation}
In the large  $N$ limit the sum over the wavevectors  ${\bf q}$ in Eq.
 (\ref{SEA}) can be converted to  an integral. For $d>8$ the integrals
 will exist with a cutoff at $q=\Lambda$, where $\Lambda \sim 1/a$ and $a$
 is the lattice spacing.   Then $\Sigma_R(0) \sim w^2 \tau^2$.  For $4<d<8$,   $\Sigma_R(0)$ does not
 require an upper cutoff and $\Sigma_R(0) \sim w^2\tau^{(d-4)/2}$. It is useful to define the coupling constant $g^2$ of the cubic theory 
\begin{equation}
g^2 =\frac{w^2}{\tau^{3-d/2}}.
\label{gdef}
\end{equation}
 Then in terms of $g^2$, $\Sigma_R(0) \sim \tau g^2$ as $g^2 \to 0$, while higher term in the loop expansion make $\Sigma_R(0) =\tau f(g^2)$ : The function $f(g^2)$  has a (weak-coupling)  series expansion  in $g^2$. Then according to Eq. (\ref{SED}) in the limit $n \to 0$ the replica symmetric solution appears to be perturbatively unstable.  To proceed further one needs the Parisi RSB scheme. However,  the extent  of replica symmetry breaking is vanishingly  small  as $g^2 \to 0$, as the breakpoint  $x_1 \sim g^2$ \cite{parisi:12}. Notice that the limit $g^2 \to 0$ is \textit{not} the critical limit, but instead is the low-temperature limit. It is possible that if the series for $f(g^2)$ could be summed to all orders the replica symmetric solution might then prove to be stable when $d < 6$. In Ref. \cite{moore:05} an example of a situation where the replica symmetric theory was found to be stable when the series was summed to all orders was explicitly constructed, even though it was unstable in low order perturbation theory as $ n \to 0$. Alas no argument for this possibility has been found for the series for $f(g^2)$  below six dimensions.

Because it cannot be demonstrated that the replica symmetric solution becomes stable when the perturbative expansion is taken to all orders, we shall adopt another approach and show that the barriers between the putative RSB pure states would be \textit{finite} if pure states existed below six dimensions. To determine barriers  one looks  for instanton solutions of the Euler-Lagrange equations associated with Eq.~(\ref{q2}) \cite{langer:69,coleman:77,mckane:79}. The instanton procedure for calculating barrier heights is usually used when there are states
which are metastable: indeed for RSB there are states with a range of free energies which differ by $O(1)$ from each other and between which one can envisage transitions. Our calculational strategy is to first calculate the leading term in the barrier height as $g^2 \to 0$, (it goes like $\sim 1/g^2$), using the replica symmetric starting point of Eq.~(\ref{q2}) and then show that RSB effects on the calculation would only add a higher order sub-dominant modification of $O(1)$. 

 We can only find instanton solutions when $d < 6$ in the (massive) breather and anomalous sectors.
The anomalous sector can be spanned by a variable $\rho_{\alpha}$ with the constraint that $\sum_{\alpha} \rho_{\alpha}=0$, while the breather sector requires just a single scalar to describe it, so that the combined breather and anomalous sector can be described by a new field $\phi_{\alpha}$, with $\alpha= 1,\ldots,n$ \cite{TDP}. In this sector the functional of Eq.~(\ref{q2}) becomes just the the effective Hamiltonian 
\begin{align}
&\tilde{H}/k_BT=\int d^dr \,\large[\sum_{\alpha=1}^n\large[\frac{1}{2} (\nabla \phi_{\alpha})^2+\frac{ 2w Q}{2} \phi_{\alpha}^2-\frac{ w}{3}\phi_{\alpha}^3 \nonumber \\
&- \sum_{\alpha \beta = 1}^n(\frac{wQ}{2} \phi_{\alpha} \phi_{\beta} + O( \phi_{\alpha}^2 \phi_{\beta} + \phi_{\alpha} \phi_{\beta}^2)) \large].
\label{Halpha}
\end{align}
If the quadratic form in $\phi_{\alpha}$ is diagonalized there are $(n-1)$ eigenvalues with mass $m_A^2$ and a single eigenvalue with mass $m_L^2$, 
The mass difference of order $n w Q$ between the breather mode and that of the anomalous mode will produce interesting effects at higher order in $g^2$ \cite{aharony:76, franz:13,biroli:14}.  We shall just look now for the instanton solutions in the $\phi_{\alpha}$ variables, and assume that it too is replica symmetric so that $\phi_{\alpha}=S$ for all $\alpha$. One can  ignore the  terms in Eq. (\ref{Halpha}) involving two replica indices as they will make a contribution of order $n^2$.  $S(\mathbf{r})$ is the instanton solution in
\begin{equation}
\tilde{H}=  nK,
\label{ndep}
\end{equation}
where
\begin{equation}
K/k_BT = \int \, d^d r\large[\frac{1}{2} (\nabla S)^2+\frac{\tau}{2} S^2-\frac{ w}{3}S^3\large],
\label{Reng}
\end{equation}
(in which the limit $n\to 0$ has been taken).

It is convenient to scale out the coefficients $\tau$ and $w$ by the variable change  
\begin{equation}
S(\mathbf{r})=\frac{\tau}{ w}P(\mathbf{r}), \hspace{0.5cm}  \mathbf{x}= \mathbf{r} \sqrt{\tau},
\label{scaling}
\end{equation}
so that distances $x$ are measured in units of the mean-field correlation length, that is,
$x = r/\xi$, where $\xi=1/\sqrt{\tau}$. Then Eq. (\ref{Reng}) can be rewritten
as
\begin{equation}
 K= \frac{\tau^{3-d/2}}{ w^2} H= \frac{1}{g^2} H,
\label{Kdef}
\end{equation}
and $H$ is given by 
\begin{equation}
H/k_BT=  \int \,d^d x\large[\frac{1}{2} (\nabla P)^2+\frac{1}{2} P^2-\frac{1}{3} P^3\large],
\label{Hdef}
\end{equation}
on setting $n=0$.
The  instanton is the spatially varying solution which makes $H$  in Eq.~(\ref{Hdef}) stationary and  is the solution of the Euler-Lagrange equation
\begin{equation}
\nabla^2 P= P -  P^2.
\label{Rstat}
\end{equation} 
 Assuming that the solution has spherical symmetry, that is  $R(\mathbf{r})=R(r)$,  the stationarity equation reduces to 
 \begin{equation}
 \frac{d^2P}{dx^2}+\frac{d-1}{x}\frac{dP}{dx}=  P-P^2.
 \label{REqm}
 \end{equation}
with boundary conditions $P(x \to \infty)\to 0$ and $dP/dx=0$ as $x \to \infty$, so that at large distance from the origin,  (which is the center of the instanton),  the replica symmetric spatially uniform mean-field solution is recovered. 

\begin{figure}
  \includegraphics[width=\columnwidth]{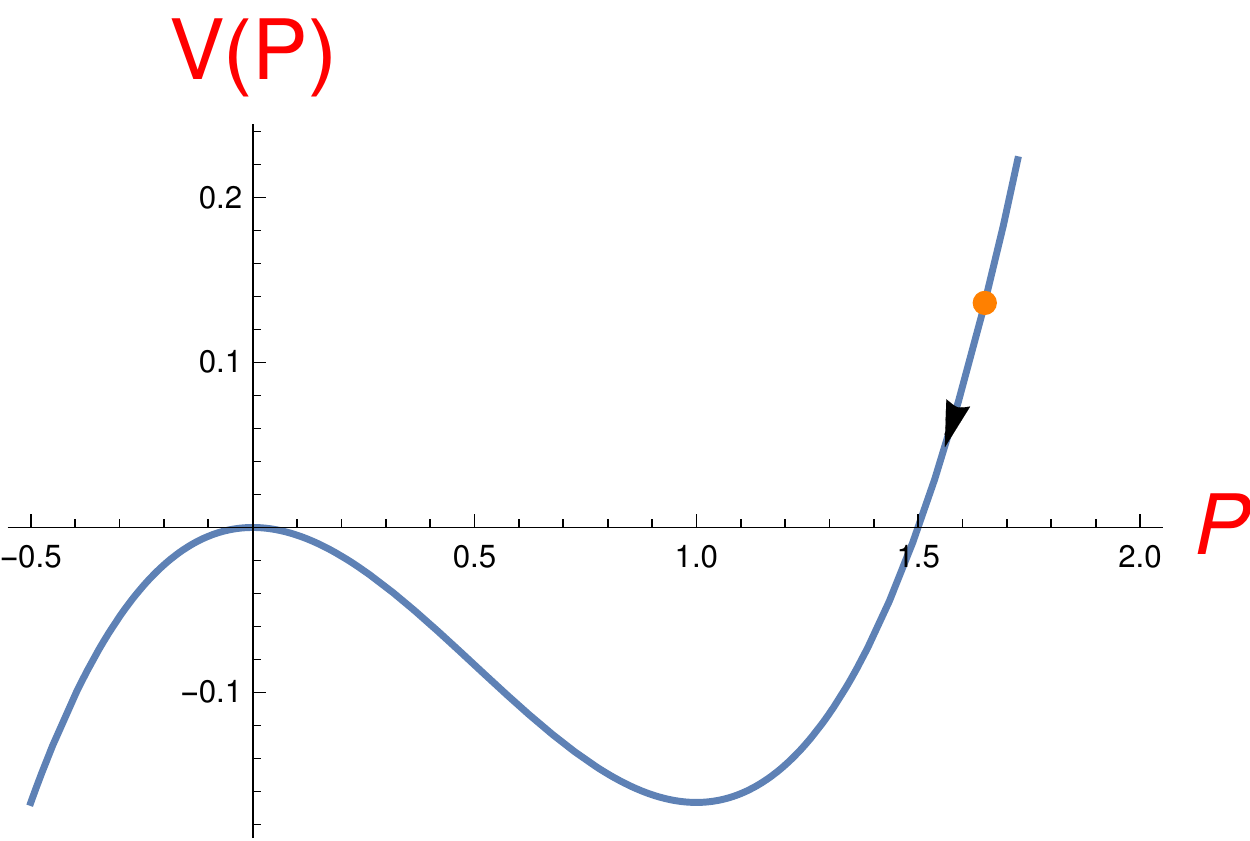}
  \caption{Plot of the potential $V(P)$ versus $P$, based upon Eq. (\ref{potential}) of the mechanical analogue. The initial point $P_0$ is marked in orange.  }
  \label{fig:mech}
\end{figure}

The solution of Eq.~(\ref{REqm}) with these boundary conditions is most easily understood by means of the \lq \lq mechanical analogue" (see Ref. \cite{coleman:77}) . The mechanical analogue consists of interpreting $P$ as a particle position and $x$ as time. The particle is moving in the potential $V[P]$ (as in Fig.~\ref{fig:mech}) which is given by
\begin{equation}
V[P]= -\frac{1}{2}  P^2+\frac{1}{3}P^3,
\label{potential}
\end{equation}
subject to a \lq \lq viscous" damping force $-\frac{d-1}{x} \, dP/dx$. Fig. \ref{fig:mech} shows that to solve the mechanical problem one has to choose the initial point on the curve, $P_0$, so that the particle can roll down the slope with sufficient speed so that it can overcome the viscous damping force and reach the origin with zero speed. This problem is readily solved numerically which is just as well as no analytical solution can  be found.    
 
The numerical solutions show that the initial value of $P$ at $x=0$,  $P_0$, has to become larger and larger as $d \to 6$ to achieve a solution: $P_0 \sim 1/(6-d)$ suggesting that $d=6$ is indeed a special dimension for the instanton solution. In fact it is readily  demonstrated \cite{unger:84,Muratov:2004,Pohozaev:65,ortega:1990}  that it is only in dimensions $d < 6$ that there exists an instanton solution of finite action.

To show this,  we first multiply both sides of Eq.~(\ref{Rstat}) by $P$ and use the identity
$\nabla \cdot( P \nabla P)= P \nabla^2P+(\nabla P)^2$ the integral over all space of $\nabla \cdot (P \nabla P)$ vanishes by Gauss's theorem and the imposed boundary conditions at $x = \infty$ so that 
\begin{equation}
\int \,d^d x\large[ (\nabla P)^2+ P^2- P^3\large]=0.
\label{identity}
\end{equation}
Using Eq. (\ref{Hdef}) one deduces that 
\begin{equation}
H/k_B T = \frac{1}{6} \int \, d^d x P^3.
\label{h1}
\end{equation} 
As $P>0$ it follows that $H>0$.

Next we obtain a second  identity  from studying dilatations of the solution $P(x)$ \cite{Pohozaev:65,ortega:1990}.  Define $P_{\lambda}(x)=\lambda^2 P(\lambda x)$ so that
\begin{align}
& H(P_{\lambda})/k_B T= \lambda^{6-d} \int \,d^d x\large[\frac{1}{2} (\nabla P)^2+\frac{1}{2 \lambda^2} P^2-\frac{1}{3} P^3\large] \nonumber \\
& = \lambda^{6-d} H(P)/k_B T+\frac{1}{2} \lambda^{6-d}(\frac{1}{\lambda^2}-1)\int \,d^d x P^2.
\label{rescale}
\end{align}
Thus
\begin{equation}
\frac{dH(P_{\lambda})}{d \lambda}=(6-d) H(P)- k_B T\int \,d^d x P^2, \hspace{0.1cm}  \textrm{at}\, \lambda=1.
\label{firstderiv}
\end{equation}
 Since $P_{\lambda}$ is a solution which makes $H$ stationary when  $\lambda =1$,
\begin{equation}
(6-d)H(P)= k_B T\int \,d^d x P^2.
\label{secondidentity}
\end{equation} 
We have already established that $H>0$ and as the integral of $P^2$ over space must be positive, this equation implies that \textit{there can only  be an  instanton solution for} $d \le 6$. 

The instanton solution should not be a stable solution of Eq. (\ref{Rstat}) as it needs to correspond to a saddle point. To study its stability 
against  a deviation $\psi(\mathbf{x})$ from the solution $P(x)$ of Eq. (\ref{Rstat}), requires solving the Schr\"{o}dinger-like eigenvalue equation,
\begin{equation}
-\nabla^2 \psi+\psi-2 P(x) \psi= \epsilon \, \psi.
\label{RSeig}
\end{equation}
Stability would require that all the eigenvalues  $\epsilon >0$. However, the lowest eigenvalue  of this equation, (the nodeless $s$-wave solution), has a negative eigenvalue, indicating that the instanton is unstable.
The negative eigenvalue actually contributes to the prefactor $\Gamma_0$ in the transition rate $\Gamma  \sim \Gamma_0 \exp(-B/k_B T)$ out of the initial   state  due to thermal fluctuations \cite{langer:69}.  An upper bound on the negative eigenvalue of $-3/(6-d)$ can be  obtained by setting $\psi \propto P(x)$.  The next eigenvalues up  are the $p$-wave modes which are $d$-fold degenerate and are null eigenvalues: they correspond to translations of the instanton. Thus there is one unstable downward  direction so the instanton solution is a  saddle point of the functional $H$ of Eq.~(\ref{Hdef}).

The instanton energy is equal to the barrier which has to be overcome to escape from the initial state. It is finite as $d \to 6$ from below and scales as $1/g^2$ according to Eq. (\ref{Kdef}). Numerical work indicates that $H/k_B T$ of Eq. (\ref{Hdef}) approaches a number $\approx 40.8$ as $d\to 6$, so  that
\begin{equation}
 B/k_B T \sim 40.8/g^2, \hspace{0.5cm} g^2 \to 0.
\label{barrier}
\end{equation}
 For $ d> 6$ there is no instanton solution of finite action (energy), so that RSB should be stable in these dimensions, while for $d < 6$ the existence of instantons of finite action implies that the barriers between pure states are finite, which would mean that RSB should not exist in these dimensions. Thermal fluctuations will cause the pure states to mix together. 

There are loop corrections in ascending powers of $g^2$ to the non-perturbative leading term for the barrier height $B$ in Eq. (\ref{barrier}), arising from several sources, including the effects of replica symmetry breaking, the coupling to the replicon fields and the differences between the breather and anomalous mode masses, and dealing with them will be hard, harder than extending the loop expansion around the spatially constant mean-field solution beyond Gaussian order, which has yet to be  done.   It is the device of studying the limit $g^2 \to 0$  which allows progress. For example, the ``box diagram'' of the cubic field theory of Eq.~(\ref{FQ}) adds to the free energy  functional an effective quartic term, $w^4 \tau^{d/2-4} Q_{\alpha \beta}^4$  \cite{green:83,fisher:85}. At mean-field level this quartic term is responsible for replica symmetry breaking \cite{parisi:79,moore:18}. It adds a contribution $g^2 P^4$ to an effective rescaled Hamiltonian as in Eq.~(\ref{Hdef}), which is negligible  compared to the terms $P^2/2 - P^3/3$ in the limit $g^2 \to 0$;  the term in $g^2 P^4$ gives a simple perturbative correction to the barrier height -- a term  of $O(1)$  on the right hand side of Eq.~(\ref{barrier}). A quartic term  which is not vanishingly small would have had important consequences, just as it does for the instantons in spinodal nucleation theory \cite{unger:84,Muratov:2004}.
 
The SK model is the $d \to \infty$ limit of spin glasses and has a  free energy landscape which is well-understood \cite{aspelmeier:04} from  solutions of the mean-field equations of Thouless, Anderson and Palmer (TAP) \cite{thouless:77}. The solutions of low free energy correspond to the pure states, and at finite $N$ there is a path from each minimum of the free energy functional to a saddle point (there is an associated saddle point for every minimum \cite{aspelmeier:04, aspelmeier:06, cavagna:04}).  The saddle point has  just one negative eigenvalue. The barrier is then the difference in free energy between the saddle point free energy and that of its associated minimum and  this  is thought to vary as $ N^{1/3}$ \cite{aspelmeier:06}.
   We believe that the replica symmetric instanton in the breather-anomalous sector provides some similarity with  the TAP landscape picture.  Thus if we restored the replica index $\alpha$ to $\psi_{\alpha}$ in Eq.~(\ref{RSeig}), there would have been $n$ identical negative eigenvalues of the functional $\tilde{H}$ of Eq.~(\ref{Halpha}), one for each of the $n$ copies in the replicated system, which  mimics the single negative eigenvalue at the saddle point for the TAP equations, (which are not replicated)  \cite{aspelmeier:04, aspelmeier:06, cavagna:04}.  
   
One difference with the barriers of the SK model is that in that model of order  $N$ spins are involved in the escape from a pure state  \cite{aspelmeier:04, aspelmeier:06, cavagna:04}, whereas in the instanton solution for $d < 6$ the changes are localized over a finite region whose size is the correlation length $\xi$ (see  Eq.~(\ref{scaling})). It follows from Eq.~(\ref{fQmin}) that the free energy density is of order $\sim \tau^3/w^2$ and so a modification of this by a spatial variation (as in the instanton) over a region of size $\xi =1/\sqrt{\tau}$ will have a total energy cost of $\sim(\tau^3/w^2) \xi^d \sim 1/g^2$. It is thus very natural that the barrier height should  be finite and vary as $1/g^2$.  But this requires the existence of an instanton solution, which is only possible for $d < 6$. The instanton solution corresponds to the critical droplet which once nucleated allows escape from the pure state. For stability a pure state needs
to be stable against the nucleation of other states within it, and that is not the case when $d<6$.

\begin{acknowledgments}

I would like to thank Alan McKane for introducing me to instantons, Niels Walet for sharing his insights  on their stability, and Giulio Biroli and Nick Read for valuable comments.

\end{acknowledgments}

\bibliography{refs}

\begin{thebibliography}{45}%
\makeatletter
\providecommand \@ifxundefined [1]{%
 \@ifx{#1\undefined}
}%
\providecommand \@ifnum [1]{%
 \ifnum #1\expandafter \@firstoftwo
 \else \expandafter \@secondoftwo
 \fi
}%
\providecommand \@ifx [1]{%
 \ifx #1\expandafter \@firstoftwo
 \else \expandafter \@secondoftwo
 \fi
}%
\providecommand \natexlab [1]{#1}%
\providecommand \enquote  [1]{``#1''}%
\providecommand \bibnamefont  [1]{#1}%
\providecommand \bibfnamefont [1]{#1}%
\providecommand \citenamefont [1]{#1}%
\providecommand \href@noop [0]{\@secondoftwo}%
\providecommand \href [0]{\begingroup \@sanitize@url \@href}%
\providecommand \@href[1]{\@@startlink{#1}\@@href}%
\providecommand \@@href[1]{\endgroup#1\@@endlink}%
\providecommand \@sanitize@url [0]{\catcode `\\12\catcode `\$12\catcode
  `\&12\catcode `\#12\catcode `\^12\catcode `\_12\catcode `\%12\relax}%
\providecommand \@@startlink[1]{}%
\providecommand \@@endlink[0]{}%
\providecommand \url  [0]{\begingroup\@sanitize@url \@url }%
\providecommand \@url [1]{\endgroup\@href {#1}{\urlprefix }}%
\providecommand \urlprefix  [0]{URL }%
\providecommand \Eprint [0]{\href }%
\providecommand \doibase [0]{http://dx.doi.org/}%
\providecommand \selectlanguage [0]{\@gobble}%
\providecommand \bibinfo  [0]{\@secondoftwo}%
\providecommand \bibfield  [0]{\@secondoftwo}%
\providecommand \translation [1]{[#1]}%
\providecommand \BibitemOpen [0]{}%
\providecommand \bibitemStop [0]{}%
\providecommand \bibitemNoStop [0]{.\EOS\space}%
\providecommand \EOS [0]{\spacefactor3000\relax}%
\providecommand \BibitemShut  [1]{\csname bibitem#1\endcsname}%
\let\auto@bib@innerbib\@empty
%</preamble>
\bibitem [{\citenamefont {Parisi}(1979)}]{parisi:79}%
  \BibitemOpen
  \bibfield  {author} {\bibinfo {author} {\bibfnamefont {G.}~\bibnamefont
  {Parisi}},\ }\bibfield  {title} {\enquote {\bibinfo {title} {Infinite number
  of order parameters for spin-glasses},}\ }\href@noop {} {\bibfield  {journal}
  {\bibinfo  {journal} {Phys. Rev. Lett.}\ }\textbf {\bibinfo {volume} {43}},\
  \bibinfo {pages} {1754} (\bibinfo {year} {1979})}\BibitemShut {NoStop}%
\bibitem [{\citenamefont {Parisi}(1983)}]{parisi:83}%
  \BibitemOpen
  \bibfield  {author} {\bibinfo {author} {\bibfnamefont {G.}~\bibnamefont
  {Parisi}},\ }\bibfield  {title} {\enquote {\bibinfo {title} {Order parameter
  for spin-glasses},}\ }\href@noop {} {\bibfield  {journal} {\bibinfo
  {journal} {Phys. Rev. Lett.}\ }\textbf {\bibinfo {volume} {50}},\ \bibinfo
  {pages} {1946} (\bibinfo {year} {1983})}\BibitemShut {NoStop}%
\bibitem [{\citenamefont {Rammal}\ \emph {et~al.}(1986)\citenamefont {Rammal},
  \citenamefont {Toulouse},\ and\ \citenamefont {Virasoro}}]{rammal:86}%
  \BibitemOpen
  \bibfield  {author} {\bibinfo {author} {\bibfnamefont {R.}~\bibnamefont
  {Rammal}}, \bibinfo {author} {\bibfnamefont {G.}~\bibnamefont {Toulouse}}, \
  and\ \bibinfo {author} {\bibfnamefont {M.~A.}\ \bibnamefont {Virasoro}},\
  }\bibfield  {title} {\enquote {\bibinfo {title} {{Ultrametricity for
  physicists}},}\ }\href@noop {} {\bibfield  {journal} {\bibinfo  {journal}
  {Rev. Mod. Phys.}\ }\textbf {\bibinfo {volume} {58}},\ \bibinfo {pages} {765}
  (\bibinfo {year} {1986})}\BibitemShut {NoStop}%
\bibitem [{\citenamefont {M\'ezard}\ \emph {et~al.}(1987)\citenamefont
  {M\'ezard}, \citenamefont {Parisi},\ and\ \citenamefont
  {Virasoro}}]{mezard:87}%
  \BibitemOpen
  \bibfield  {author} {\bibinfo {author} {\bibfnamefont {M.}~\bibnamefont
  {M\'ezard}}, \bibinfo {author} {\bibfnamefont {G.}~\bibnamefont {Parisi}}, \
  and\ \bibinfo {author} {\bibfnamefont {M.~A.}\ \bibnamefont {Virasoro}},\
  }\href@noop {} {\emph {\bibinfo {title} {Spin Glass Theory and Beyond}}}\
  (\bibinfo  {publisher} {World Scientific},\ \bibinfo {address} {Singapore},\
  \bibinfo {year} {1987})\BibitemShut {NoStop}%
\bibitem [{\citenamefont {Parisi}(2008)}]{parisi:08}%
  \BibitemOpen
  \bibfield  {author} {\bibinfo {author} {\bibfnamefont {G.}~\bibnamefont
  {Parisi}},\ }\bibfield  {title} {\enquote {\bibinfo {title} {{{Some
  considerations of finite dimensional spin glasses}}},}\ }\href@noop {}
  {\bibfield  {journal} {\bibinfo  {journal} {J. Phys. A}\ }\textbf {\bibinfo
  {volume} {41}},\ \bibinfo {pages} {324002} (\bibinfo {year}
  {2008})}\BibitemShut {NoStop}%
\bibitem [{\citenamefont {McMillan}(1984)}]{mcmillan:84a}%
  \BibitemOpen
  \bibfield  {author} {\bibinfo {author} {\bibfnamefont {W.~L.}\ \bibnamefont
  {McMillan}},\ }\bibfield  {title} {\enquote {\bibinfo {title} {{Scaling
  theory of {I}sing spin glasses}},}\ }\href@noop {} {\bibfield  {journal}
  {\bibinfo  {journal} {J. Phys. C}\ }\textbf {\bibinfo {volume} {17}},\
  \bibinfo {pages} {3179} (\bibinfo {year} {1984})}\BibitemShut {NoStop}%
\bibitem [{\citenamefont {Bray}\ and\ \citenamefont {Moore}(1986)}]{bray:86}%
  \BibitemOpen
  \bibfield  {author} {\bibinfo {author} {\bibfnamefont {A.~J.}\ \bibnamefont
  {Bray}}\ and\ \bibinfo {author} {\bibfnamefont {M.~A.}\ \bibnamefont
  {Moore}},\ }\bibfield  {title} {\enquote {\bibinfo {title} {Scaling theory of
  the ordered phase of spin glasses},}\ }in\ \href@noop {} {\emph {\bibinfo
  {booktitle} {{Heidelberg Colloquium on Glassy Dynamics and Optimization}}}},\
  \bibinfo {editor} {edited by\ \bibinfo {editor} {\bibfnamefont
  {L.}~\bibnamefont {Van~Hemmen}}\ and\ \bibinfo {editor} {\bibfnamefont
  {I.}~\bibnamefont {Morgenstern}}}\ (\bibinfo  {publisher} {Springer},\
  \bibinfo {address} {New York},\ \bibinfo {year} {1986})\ p.\ \bibinfo {pages}
  {121}\BibitemShut {NoStop}%
\bibitem [{\citenamefont {Fisher}\ and\ \citenamefont
  {Huse}(1988)}]{fisher:88}%
  \BibitemOpen
  \bibfield  {author} {\bibinfo {author} {\bibfnamefont {D.~S.}\ \bibnamefont
  {Fisher}}\ and\ \bibinfo {author} {\bibfnamefont {D.~A.}\ \bibnamefont
  {Huse}},\ }\bibfield  {title} {\enquote {\bibinfo {title} {{Equilibrium
  behavior of the spin-glass ordered phase}},}\ }\href@noop {} {\bibfield
  {journal} {\bibinfo  {journal} {Phys. Rev. B}\ }\textbf {\bibinfo {volume}
  {38}},\ \bibinfo {pages} {386} (\bibinfo {year} {1988})}\BibitemShut
  {NoStop}%
\bibitem [{\citenamefont {Moore}\ \emph {et~al.}(1998)\citenamefont {Moore},
  \citenamefont {Bokil},\ and\ \citenamefont {Drossel}}]{moore:98}%
  \BibitemOpen
  \bibfield  {author} {\bibinfo {author} {\bibfnamefont {M.~A.}\ \bibnamefont
  {Moore}}, \bibinfo {author} {\bibfnamefont {Hemant}\ \bibnamefont {Bokil}}, \
  and\ \bibinfo {author} {\bibfnamefont {Barbara}\ \bibnamefont {Drossel}},\
  }\bibfield  {title} {\enquote {\bibinfo {title} {Evidence for the droplet
  picture of spin glasses},}\ }\href@noop {} {\bibfield  {journal} {\bibinfo
  {journal} {Phys. Rev. Lett.}\ }\textbf {\bibinfo {volume} {81}},\ \bibinfo
  {pages} {4252} (\bibinfo {year} {1998})}\BibitemShut {NoStop}%
\bibitem [{\citenamefont {Wang}\ \emph {et~al.}(2017)\citenamefont {Wang},
  \citenamefont {Moore},\ and\ \citenamefont {Katzgraber}}]{wang:17}%
  \BibitemOpen
  \bibfield  {author} {\bibinfo {author} {\bibfnamefont {Wenlong}\ \bibnamefont
  {Wang}}, \bibinfo {author} {\bibfnamefont {M.~A.}\ \bibnamefont {Moore}}, \
  and\ \bibinfo {author} {\bibfnamefont {Helmut~G.}\ \bibnamefont
  {Katzgraber}},\ }\bibfield  {title} {\enquote {\bibinfo {title} {{Fractal
  Dimension of Interfaces in Edwards-Anderson and Long-range Ising Spin
  Glasses: Determining the Applicability of Different Theoretical
  Descriptions}},}\ }\href@noop {} {\bibfield  {journal} {\bibinfo  {journal}
  {Phys. Rev. Lett.}\ }\textbf {\bibinfo {volume} {119}},\ \bibinfo {pages}
  {100602} (\bibinfo {year} {2017})}\BibitemShut {NoStop}%
\bibitem [{\citenamefont {{Angelini}}\ and\ \citenamefont
  {{Biroli}}(2015)}]{angelini:15}%
  \BibitemOpen
  \bibfield  {author} {\bibinfo {author} {\bibfnamefont {M.~C.}\ \bibnamefont
  {{Angelini}}}\ and\ \bibinfo {author} {\bibfnamefont {G.}~\bibnamefont
  {{Biroli}}},\ }\bibfield  {title} {\enquote {\bibinfo {title} {{{Spin Glass
  in a Field: A New Zero-Temperature Fixed Point in Finite Dimensions}}},}\
  }\href@noop {} {\bibfield  {journal} {\bibinfo  {journal} {Phys. Rev. Lett.}\
  }\textbf {\bibinfo {volume} {114}},\ \bibinfo {pages} {095701} (\bibinfo
  {year} {2015})}\BibitemShut {NoStop}%
\bibitem [{\citenamefont {{Angelini}}\ and\ \citenamefont
  {{Biroli}}(2017)}]{angelini:17}%
  \BibitemOpen
  \bibfield  {author} {\bibinfo {author} {\bibfnamefont {M.~C.}\ \bibnamefont
  {{Angelini}}}\ and\ \bibinfo {author} {\bibfnamefont {G.}~\bibnamefont
  {{Biroli}}},\ }\bibfield  {title} {\enquote {\bibinfo {title} {{{Real space
  renormalization group of disordered models of glasses}}},}\ }\href@noop {}
  {\bibfield  {journal} {\bibinfo  {journal} {Proc. Natl. Acad. Sci. U.~S.~A.}\
  }\textbf {\bibinfo {volume} {114}},\ \bibinfo {pages} {3328} (\bibinfo {year}
  {2017})}\BibitemShut {NoStop}%
\bibitem [{\citenamefont {Angelini}\ and\ \citenamefont
  {Biroli}(2017)}]{angelini:17a}%
  \BibitemOpen
  \bibfield  {author} {\bibinfo {author} {\bibfnamefont {Maria~Chiara}\
  \bibnamefont {Angelini}}\ and\ \bibinfo {author} {\bibfnamefont {Giulio}\
  \bibnamefont {Biroli}},\ }\bibfield  {title} {\enquote {\bibinfo {title}
  {{{Real Space Migdal--Kadanoff Renormalisation of Glassy Systems: Recent
  Results and a Critical Assessment}}},}\ }\href@noop {} {\bibfield  {journal}
  {\bibinfo  {journal} {Journal of Statistical Physics}\ }\textbf {\bibinfo
  {volume} {167}},\ \bibinfo {pages} {476} (\bibinfo {year}
  {2017})}\BibitemShut {NoStop}%
\bibitem [{\citenamefont {Wang}\ \emph {et~al.}(2018)\citenamefont {Wang},
  \citenamefont {Moore},\ and\ \citenamefont {Katzgraber}}]{wang:18b}%
  \BibitemOpen
  \bibfield  {author} {\bibinfo {author} {\bibfnamefont {Wenlong}\ \bibnamefont
  {Wang}}, \bibinfo {author} {\bibfnamefont {M.~A.}\ \bibnamefont {Moore}}, \
  and\ \bibinfo {author} {\bibfnamefont {Helmut~G.}\ \bibnamefont
  {Katzgraber}},\ }\bibfield  {title} {\enquote {\bibinfo {title} {{Fractal
  dimension of interfaces in Edwards-Anderson spin glasses for up to six space
  dimensions}},}\ }\href@noop {} {\bibfield  {journal} {\bibinfo  {journal}
  {Phys. Rev. E}\ }\textbf {\bibinfo {volume} {97}},\ \bibinfo {pages} {032104}
  (\bibinfo {year} {2018})}\BibitemShut {NoStop}%
\bibitem [{\citenamefont {Moore}\ and\ \citenamefont {Bray}(2011)}]{Moore:11}%
  \BibitemOpen
  \bibfield  {author} {\bibinfo {author} {\bibfnamefont {M.~A.}\ \bibnamefont
  {Moore}}\ and\ \bibinfo {author} {\bibfnamefont {A.~J.}\ \bibnamefont
  {Bray}},\ }\bibfield  {title} {\enquote {\bibinfo {title} {{{Disappearance of
  the de Almeida-Thouless line in six dimensions}}},}\ }\href@noop {}
  {\bibfield  {journal} {\bibinfo  {journal} {Phys. Rev. B}\ }\textbf {\bibinfo
  {volume} {83}},\ \bibinfo {pages} {224408} (\bibinfo {year}
  {2011})}\BibitemShut {NoStop}%
\bibitem [{\citenamefont {Moore}\ and\ \citenamefont {Read}(2018)}]{moore:18}%
  \BibitemOpen
  \bibfield  {author} {\bibinfo {author} {\bibfnamefont {M.~A.}\ \bibnamefont
  {Moore}}\ and\ \bibinfo {author} {\bibfnamefont {N.}~\bibnamefont {Read}},\
  }\bibfield  {title} {\enquote {\bibinfo {title} {{Multicritical Point on the
  de Almeida--Thouless Line in Spin Glasses in $d > 6$ Dimensions}},}\
  }\href@noop {} {\bibfield  {journal} {\bibinfo  {journal} {Phys. Rev. Lett.}\
  }\textbf {\bibinfo {volume} {120}},\ \bibinfo {pages} {130602} (\bibinfo
  {year} {2018})}\BibitemShut {NoStop}%
\bibitem [{\citenamefont {Parisi}\ and\ \citenamefont
  {Temesv\'{a}ri}(2012)}]{parisi:12}%
  \BibitemOpen
  \bibfield  {author} {\bibinfo {author} {\bibfnamefont {G.}~\bibnamefont
  {Parisi}}\ and\ \bibinfo {author} {\bibfnamefont {T.}~\bibnamefont
  {Temesv\'{a}ri}},\ }\bibfield  {title} {\enquote {\bibinfo {title} {Replica
  symmetry breaking in and around six dimensions},}\ }\href@noop {} {\bibfield
  {journal} {\bibinfo  {journal} {Nuclear Physics B}\ }\textbf {\bibinfo
  {volume} {858}},\ \bibinfo {pages} {293} (\bibinfo {year}
  {2012})}\BibitemShut {NoStop}%
\bibitem [{\citenamefont {Temesv\'ari}(2017)}]{tamas:17}%
  \BibitemOpen
  \bibfield  {author} {\bibinfo {author} {\bibfnamefont {T.}~\bibnamefont
  {Temesv\'ari}},\ }\bibfield  {title} {\enquote {\bibinfo {title} {{Physical
  observables of the Ising spin glass in $6\ensuremath{-}\ensuremath{\epsilon}$
  dimensions: Asymptotical behavior around the critical fixed point}},}\
  }\href@noop {} {\bibfield  {journal} {\bibinfo  {journal} {Phys. Rev. B}\
  }\textbf {\bibinfo {volume} {96}},\ \bibinfo {pages} {024411} (\bibinfo
  {year} {2017})}\BibitemShut {NoStop}%
\bibitem [{\citenamefont {Charbonneau}\ and\ \citenamefont
  {Yaida}(2017)}]{yaida:17}%
  \BibitemOpen
  \bibfield  {author} {\bibinfo {author} {\bibfnamefont {Patrick}\ \bibnamefont
  {Charbonneau}}\ and\ \bibinfo {author} {\bibfnamefont {Sho}\ \bibnamefont
  {Yaida}},\ }\bibfield  {title} {\enquote {\bibinfo {title} {{Nontrivial
  Critical Fixed Point for Replica-Symmetry-Breaking Transitions}},}\
  }\href@noop {} {\bibfield  {journal} {\bibinfo  {journal} {Phys. Rev. Lett.}\
  }\textbf {\bibinfo {volume} {118}},\ \bibinfo {pages} {215701} (\bibinfo
  {year} {2017})}\BibitemShut {NoStop}%
\bibitem [{\citenamefont {Billoire}(2010)}]{Billoire:10}%
  \BibitemOpen
  \bibfield  {author} {\bibinfo {author} {\bibfnamefont {Alain}\ \bibnamefont
  {Billoire}},\ }\bibfield  {title} {\enquote {\bibinfo {title} {{Distribution
  of timescales in the Sherrington{\textendash}Kirkpatrick model}},}\
  }\href@noop {} {\bibfield  {journal} {\bibinfo  {journal} {Journal of
  Statistical Mechanics: Theory and Experiment}\ }\textbf {\bibinfo {volume}
  {2010}},\ \bibinfo {pages} {P11034} (\bibinfo {year} {2010})}\BibitemShut
  {NoStop}%
\bibitem [{\citenamefont {Billoire}\ and\ \citenamefont
  {Marinari}(2001)}]{billoire:01}%
  \BibitemOpen
  \bibfield  {author} {\bibinfo {author} {\bibfnamefont {Alain}\ \bibnamefont
  {Billoire}}\ and\ \bibinfo {author} {\bibfnamefont {Enzo}\ \bibnamefont
  {Marinari}},\ }\bibfield  {title} {\enquote {\bibinfo {title} {{Correlation
  timescales in the Sherrington-Kirkpatrick model}},}\ }\href@noop {}
  {\bibfield  {journal} {\bibinfo  {journal} {Journal of Physics A:
  Mathematical and General}\ }\textbf {\bibinfo {volume} {34}},\ \bibinfo
  {pages} {L727--L734} (\bibinfo {year} {2001})}\BibitemShut {NoStop}%
\bibitem [{\citenamefont {Sherrington}\ and\ \citenamefont
  {Kirkpatrick}(1975)}]{sherrington:75}%
  \BibitemOpen
  \bibfield  {author} {\bibinfo {author} {\bibfnamefont {David}\ \bibnamefont
  {Sherrington}}\ and\ \bibinfo {author} {\bibfnamefont {Scott}\ \bibnamefont
  {Kirkpatrick}},\ }\bibfield  {title} {\enquote {\bibinfo {title} {{Solvable
  Model of a Spin-Glass}},}\ }\href@noop {} {\bibfield  {journal} {\bibinfo
  {journal} {Phys. Rev. Lett.}\ }\textbf {\bibinfo {volume} {35}},\ \bibinfo
  {pages} {1792} (\bibinfo {year} {1975})}\BibitemShut {NoStop}%
\bibitem [{\citenamefont {Rodgers}\ and\ \citenamefont
  {Moore}(1989)}]{Rodgers:89}%
  \BibitemOpen
  \bibfield  {author} {\bibinfo {author} {\bibfnamefont {G~J}\ \bibnamefont
  {Rodgers}}\ and\ \bibinfo {author} {\bibfnamefont {M~A}\ \bibnamefont
  {Moore}},\ }\bibfield  {title} {\enquote {\bibinfo {title} {{Distribution of
  barrier heights in infinite-range spin glass models}},}\ }\href@noop {}
  {\bibfield  {journal} {\bibinfo  {journal} {Journal of Physics A:
  Mathematical and General}\ }\textbf {\bibinfo {volume} {22}},\ \bibinfo
  {pages} {1085--1100} (\bibinfo {year} {1989})}\BibitemShut {NoStop}%
\bibitem [{\citenamefont {Aspelmeier}\ \emph {et~al.}(2006)\citenamefont
  {Aspelmeier}, \citenamefont {Blythe}, \citenamefont {Bray},\ and\
  \citenamefont {Moore}}]{aspelmeier:06}%
  \BibitemOpen
  \bibfield  {author} {\bibinfo {author} {\bibfnamefont {T.}~\bibnamefont
  {Aspelmeier}}, \bibinfo {author} {\bibfnamefont {R.~A.}\ \bibnamefont
  {Blythe}}, \bibinfo {author} {\bibfnamefont {A.~J.}\ \bibnamefont {Bray}}, \
  and\ \bibinfo {author} {\bibfnamefont {M.~A.}\ \bibnamefont {Moore}},\
  }\bibfield  {title} {\enquote {\bibinfo {title} {{Free-energy landscapes,
  dynamics, and the edge of chaos in mean-field models of spin glasses}},}\
  }\href@noop {} {\bibfield  {journal} {\bibinfo  {journal} {Phys. Rev. B}\
  }\textbf {\bibinfo {volume} {74}},\ \bibinfo {pages} {184411} (\bibinfo
  {year} {2006})}\BibitemShut {NoStop}%
\bibitem [{\citenamefont {Edwards}\ and\ \citenamefont
  {Anderson}(1975)}]{edwards:75}%
  \BibitemOpen
  \bibfield  {author} {\bibinfo {author} {\bibfnamefont {S.~F.}\ \bibnamefont
  {Edwards}}\ and\ \bibinfo {author} {\bibfnamefont {P.~W.}\ \bibnamefont
  {Anderson}},\ }\bibfield  {title} {\enquote {\bibinfo {title} {Theory of spin
  glasses},}\ }\href@noop {} {\bibfield  {journal} {\bibinfo  {journal} {J.
  Phys. F: Met. Phys.}\ }\textbf {\bibinfo {volume} {5}},\ \bibinfo {pages}
  {965} (\bibinfo {year} {1975})}\BibitemShut {NoStop}%
\bibitem [{\citenamefont {Green}\ \emph {et~al.}(1983)\citenamefont {Green},
  \citenamefont {Moore},\ and\ \citenamefont {Bray}}]{green:83}%
  \BibitemOpen
  \bibfield  {author} {\bibinfo {author} {\bibfnamefont {J.~E.}\ \bibnamefont
  {Green}}, \bibinfo {author} {\bibfnamefont {M.~A.}\ \bibnamefont {Moore}}, \
  and\ \bibinfo {author} {\bibfnamefont {A.~J.}\ \bibnamefont {Bray}},\
  }\bibfield  {title} {\enquote {\bibinfo {title} {{Upper critical dimension
  for the de Almeida-Thouless instability in spin glasses}},}\ }\href@noop {}
  {\bibfield  {journal} {\bibinfo  {journal} {Journal of Physics C: Solid State
  Physics}\ }\textbf {\bibinfo {volume} {16}},\ \bibinfo {pages} {L815}
  (\bibinfo {year} {1983})}\BibitemShut {NoStop}%
\bibitem [{\citenamefont {Harris}\ \emph {et~al.}(1976)\citenamefont {Harris},
  \citenamefont {Lubensky},\ and\ \citenamefont {Chen}}]{harris:76}%
  \BibitemOpen
  \bibfield  {author} {\bibinfo {author} {\bibfnamefont {A.~B.}\ \bibnamefont
  {Harris}}, \bibinfo {author} {\bibfnamefont {T.~C.}\ \bibnamefont
  {Lubensky}}, \ and\ \bibinfo {author} {\bibfnamefont {Jing-Huei}\
  \bibnamefont {Chen}},\ }\bibfield  {title} {\enquote {\bibinfo {title}
  {{Critical Properties of Spin-Glasses}},}\ }\href@noop {} {\bibfield
  {journal} {\bibinfo  {journal} {Phys. Rev. Lett.}\ }\textbf {\bibinfo
  {volume} {36}},\ \bibinfo {pages} {415} (\bibinfo {year} {1976})}\BibitemShut
  {NoStop}%
\bibitem [{\citenamefont {Pytte}\ and\ \citenamefont
  {Rudnick}(1979)}]{pytte:79}%
  \BibitemOpen
  \bibfield  {author} {\bibinfo {author} {\bibfnamefont {E.}~\bibnamefont
  {Pytte}}\ and\ \bibinfo {author} {\bibfnamefont {Joseph}\ \bibnamefont
  {Rudnick}},\ }\bibfield  {title} {\enquote {\bibinfo {title} {{Scaling,
  equation of state, and the instability of the spin-glass phase}},}\
  }\href@noop {} {\bibfield  {journal} {\bibinfo  {journal} {Phys. Rev. B}\
  }\textbf {\bibinfo {volume} {19}},\ \bibinfo {pages} {3603} (\bibinfo {year}
  {1979})}\BibitemShut {NoStop}%
\bibitem [{\citenamefont {Bray}\ and\ \citenamefont {Moore}(1979)}]{bray:79}%
  \BibitemOpen
  \bibfield  {author} {\bibinfo {author} {\bibfnamefont {A.~J.}\ \bibnamefont
  {Bray}}\ and\ \bibinfo {author} {\bibfnamefont {M.~A.}\ \bibnamefont
  {Moore}},\ }\bibfield  {title} {\enquote {\bibinfo {title} {{Replica symmetry
  and massless modes in the Ising spin glass}},}\ }\href@noop {} {\bibfield
  {journal} {\bibinfo  {journal} {Journal of Physics C: Solid State Physics}\
  }\textbf {\bibinfo {volume} {12}},\ \bibinfo {pages} {79} (\bibinfo {year}
  {1979})}\BibitemShut {NoStop}%
\bibitem [{\citenamefont {Temesv{\'a}ri}\ \emph {et~al.}(2002)\citenamefont
  {Temesv{\'a}ri}, \citenamefont {De~Dominicis},\ and\ \citenamefont
  {Pimentel}}]{TDP}%
  \BibitemOpen
  \bibfield  {author} {\bibinfo {author} {\bibfnamefont {T.}~\bibnamefont
  {Temesv{\'a}ri}}, \bibinfo {author} {\bibfnamefont {C.}~\bibnamefont
  {De~Dominicis}}, \ and\ \bibinfo {author} {\bibfnamefont {I.R.}\ \bibnamefont
  {Pimentel}},\ }\bibfield  {title} {\enquote {\bibinfo {title} {{{Generic
  replica symmetric field-theory for short range Ising spin glasses}}},}\
  }\href@noop {} {\bibfield  {journal} {\bibinfo  {journal} {The European
  Physical Journal B - Condensed Matter and Complex Systems}\ }\textbf
  {\bibinfo {volume} {25}},\ \bibinfo {pages} {361} (\bibinfo {year}
  {2002})}\BibitemShut {NoStop}%
\bibitem [{\citenamefont {Moore}(2005)}]{moore:05}%
  \BibitemOpen
  \bibfield  {author} {\bibinfo {author} {\bibfnamefont {M~A}\ \bibnamefont
  {Moore}},\ }\bibfield  {title} {\enquote {\bibinfo {title} {{The stability of
  the replica-symmetric state in finite-dimensional spin glasses}},}\
  }\href@noop {} {\bibfield  {journal} {\bibinfo  {journal} {Journal of Physics
  A: Mathematical and General}\ }\textbf {\bibinfo {volume} {38}},\ \bibinfo
  {pages} {L783--L789} (\bibinfo {year} {2005})}\BibitemShut {NoStop}%
\bibitem [{\citenamefont {Langer}(1969)}]{langer:69}%
  \BibitemOpen
  \bibfield  {author} {\bibinfo {author} {\bibfnamefont {J.S.}\ \bibnamefont
  {Langer}},\ }\bibfield  {title} {\enquote {\bibinfo {title} {Statistical
  theory of the decay of metastable states},}\ }\href@noop {} {\bibfield
  {journal} {\bibinfo  {journal} {Annals of Physics}\ }\textbf {\bibinfo
  {volume} {54}},\ \bibinfo {pages} {258} (\bibinfo {year} {1969})}\BibitemShut
  {NoStop}%
\bibitem [{\citenamefont {Coleman}(1977)}]{coleman:77}%
  \BibitemOpen
  \bibfield  {author} {\bibinfo {author} {\bibfnamefont {Sidney}\ \bibnamefont
  {Coleman}},\ }\bibfield  {title} {\enquote {\bibinfo {title} {Fate of the
  false vacuum: Semiclassical theory},}\ }\href@noop {} {\bibfield  {journal}
  {\bibinfo  {journal} {Phys. Rev. D}\ }\textbf {\bibinfo {volume} {15}},\
  \bibinfo {pages} {2929--2936} (\bibinfo {year} {1977})}\BibitemShut {NoStop}%
\bibitem [{\citenamefont {McKane}(1979)}]{mckane:79}%
  \BibitemOpen
  \bibfield  {author} {\bibinfo {author} {\bibfnamefont {A.J.}\ \bibnamefont
  {McKane}},\ }\bibfield  {title} {\enquote {\bibinfo {title} {Vacuum
  instability in scalar field theories},}\ }\href@noop {} {\bibfield  {journal}
  {\bibinfo  {journal} {Nuclear Physics B}\ }\textbf {\bibinfo {volume}
  {152}},\ \bibinfo {pages} {166 -- 188} (\bibinfo {year} {1979})}\BibitemShut
  {NoStop}%
\bibitem [{\citenamefont {Aharony}\ \emph {et~al.}(1976)\citenamefont
  {Aharony}, \citenamefont {Imry},\ and\ \citenamefont {Ma}}]{aharony:76}%
  \BibitemOpen
  \bibfield  {author} {\bibinfo {author} {\bibfnamefont {Amnon}\ \bibnamefont
  {Aharony}}, \bibinfo {author} {\bibfnamefont {Yoseph}\ \bibnamefont {Imry}},
  \ and\ \bibinfo {author} {\bibfnamefont {Shang-keng}\ \bibnamefont {Ma}},\
  }\bibfield  {title} {\enquote {\bibinfo {title} {{Lowering of Dimensionality
  in Phase Transitions with Random Fields}},}\ }\href@noop {} {\bibfield
  {journal} {\bibinfo  {journal} {Phys. Rev. Lett.}\ }\textbf {\bibinfo
  {volume} {37}},\ \bibinfo {pages} {1364--1367} (\bibinfo {year}
  {1976})}\BibitemShut {NoStop}%
\bibitem [{\citenamefont {Franz}\ \emph {et~al.}(2013)\citenamefont {Franz},
  \citenamefont {Parisi},\ and\ \citenamefont {Ricci-Tersenghi}}]{franz:13}%
  \BibitemOpen
  \bibfield  {author} {\bibinfo {author} {\bibfnamefont {Silvio}\ \bibnamefont
  {Franz}}, \bibinfo {author} {\bibfnamefont {Giorgio}\ \bibnamefont {Parisi}},
  \ and\ \bibinfo {author} {\bibfnamefont {Federico}\ \bibnamefont
  {Ricci-Tersenghi}},\ }\bibfield  {title} {\enquote {\bibinfo {title} {{Glassy
  critical points and the random field Ising model}},}\ }\href@noop {}
  {\bibfield  {journal} {\bibinfo  {journal} {Journal of Statistical Mechanics:
  Theory and Experiment}\ }\textbf {\bibinfo {volume} {2013}},\ \bibinfo
  {pages} {L02001} (\bibinfo {year} {2013})}\BibitemShut {NoStop}%
\bibitem [{\citenamefont {Biroli}\ \emph {et~al.}(2014)\citenamefont {Biroli},
  \citenamefont {Cammarota}, \citenamefont {Tarjus},\ and\ \citenamefont
  {Tarzia}}]{biroli:14}%
  \BibitemOpen
  \bibfield  {author} {\bibinfo {author} {\bibfnamefont {Giulio}\ \bibnamefont
  {Biroli}}, \bibinfo {author} {\bibfnamefont {Chiara}\ \bibnamefont
  {Cammarota}}, \bibinfo {author} {\bibfnamefont {Gilles}\ \bibnamefont
  {Tarjus}}, \ and\ \bibinfo {author} {\bibfnamefont {Marco}\ \bibnamefont
  {Tarzia}},\ }\bibfield  {title} {\enquote {\bibinfo {title}
  {Random-field-like criticality in glass-forming liquids},}\ }\href@noop {}
  {\bibfield  {journal} {\bibinfo  {journal} {Phys. Rev. Lett.}\ }\textbf
  {\bibinfo {volume} {112}},\ \bibinfo {pages} {175701} (\bibinfo {year}
  {2014})}\BibitemShut {NoStop}%
\bibitem [{\citenamefont {Unger}\ and\ \citenamefont {Klein}(1984)}]{unger:84}%
  \BibitemOpen
  \bibfield  {author} {\bibinfo {author} {\bibfnamefont {Chris}\ \bibnamefont
  {Unger}}\ and\ \bibinfo {author} {\bibfnamefont {W.}~\bibnamefont {Klein}},\
  }\bibfield  {title} {\enquote {\bibinfo {title} {{Nucleation theory near the
  classical spinodal}},}\ }\href@noop {} {\bibfield  {journal} {\bibinfo
  {journal} {Phys. Rev. B}\ }\textbf {\bibinfo {volume} {29}},\ \bibinfo
  {pages} {2698--2708} (\bibinfo {year} {1984})}\BibitemShut {NoStop}%
\bibitem [{\citenamefont {Muratov}\ and\ \citenamefont
  {Vanden-Eijnden}(2004)}]{Muratov:2004}%
  \BibitemOpen
  \bibfield  {author} {\bibinfo {author} {\bibfnamefont {Cyrill~B.}\
  \bibnamefont {Muratov}}\ and\ \bibinfo {author} {\bibfnamefont {Eric}\
  \bibnamefont {Vanden-Eijnden}},\ }\bibfield  {title} {\enquote {\bibinfo
  {title} {Breakup of universality in the generalized spinodal nucleation
  theory},}\ }\href@noop {} {\bibfield  {journal} {\bibinfo  {journal} {Journal
  of Statistical Physics}\ }\textbf {\bibinfo {volume} {114}},\ \bibinfo
  {pages} {605--623} (\bibinfo {year} {2004})}\BibitemShut {NoStop}%
\bibitem [{\citenamefont {I.}(1965)}]{Pohozaev:65}%
  \BibitemOpen
  \bibfield  {author} {\bibinfo {author} {\bibfnamefont {Pohozaev~S.}\
  \bibnamefont {I.}},\ }\bibfield  {title} {\enquote {\bibinfo {title}
  {{Eigenfunctions of the equation {$\Delta u + \lambda f(u) = 0$}}},}\
  }\href@noop {} {\bibfield  {journal} {\bibinfo  {journal} {Soviet Math.
  Doklady}\ }\textbf {\bibinfo {volume} {6}},\ \bibinfo {pages} {1408}
  (\bibinfo {year} {1965})}\BibitemShut {NoStop}%
\bibitem [{\citenamefont {Ortega}(1990)}]{ortega:1990}%
  \BibitemOpen
  \bibfield  {author} {\bibinfo {author} {\bibfnamefont {Rafael}\ \bibnamefont
  {Ortega}},\ }\bibfield  {title} {\enquote {\bibinfo {title} {{Nonexistence of
  radial solutions of two elliptic boundary value problems}},}\ }\href@noop {}
  {\bibfield  {journal} {\bibinfo  {journal} {Proceedings of the Royal Society
  of Edinburgh: Section A Mathematics}\ }\textbf {\bibinfo {volume} {114}},\
  \bibinfo {pages} {27–31} (\bibinfo {year} {1990})}\BibitemShut {NoStop}%
\bibitem [{\citenamefont {Fisher}\ and\ \citenamefont
  {Sompolinsky}(1985)}]{fisher:85}%
  \BibitemOpen
  \bibfield  {author} {\bibinfo {author} {\bibfnamefont {Daniel~S.}\
  \bibnamefont {Fisher}}\ and\ \bibinfo {author} {\bibfnamefont
  {H.}~\bibnamefont {Sompolinsky}},\ }\bibfield  {title} {\enquote {\bibinfo
  {title} {Scaling in spin-glasses},}\ }\href@noop {} {\bibfield  {journal}
  {\bibinfo  {journal} {Phys. Rev. Lett.}\ }\textbf {\bibinfo {volume} {54}},\
  \bibinfo {pages} {1063--1066} (\bibinfo {year} {1985})}\BibitemShut {NoStop}%
\bibitem [{\citenamefont {Aspelmeier}\ \emph {et~al.}(2004)\citenamefont
  {Aspelmeier}, \citenamefont {Bray},\ and\ \citenamefont
  {Moore}}]{aspelmeier:04}%
  \BibitemOpen
  \bibfield  {author} {\bibinfo {author} {\bibfnamefont {T.}~\bibnamefont
  {Aspelmeier}}, \bibinfo {author} {\bibfnamefont {A.~J.}\ \bibnamefont
  {Bray}}, \ and\ \bibinfo {author} {\bibfnamefont {M.~A.}\ \bibnamefont
  {Moore}},\ }\bibfield  {title} {\enquote {\bibinfo {title} {{Complexity of
  Ising Spin Glasses}},}\ }\href@noop {} {\bibfield  {journal} {\bibinfo
  {journal} {Phys. Rev. Lett.}\ }\textbf {\bibinfo {volume} {92}},\ \bibinfo
  {pages} {087203} (\bibinfo {year} {2004})}\BibitemShut {NoStop}%
\bibitem [{\citenamefont {Thouless}\ \emph {et~al.}(1977)\citenamefont
  {Thouless}, \citenamefont {Anderson},\ and\ \citenamefont
  {Palmer}}]{thouless:77}%
  \BibitemOpen
  \bibfield  {author} {\bibinfo {author} {\bibfnamefont {D.~J.}\ \bibnamefont
  {Thouless}}, \bibinfo {author} {\bibfnamefont {P.~W.}\ \bibnamefont
  {Anderson}}, \ and\ \bibinfo {author} {\bibfnamefont {R.~G.}\ \bibnamefont
  {Palmer}},\ }\bibfield  {title} {\enquote {\bibinfo {title} {{Solution of
  'Solvable model of a spin glass'}},}\ }\href@noop {} {\bibfield  {journal}
  {\bibinfo  {journal} {The Philosophical Magazine: A Journal of Theoretical
  Experimental and Applied Physics}\ }\textbf {\bibinfo {volume} {35}},\
  \bibinfo {pages} {593--601} (\bibinfo {year} {1977})}\BibitemShut {NoStop}%
\bibitem [{\citenamefont {Cavagna}\ \emph {et~al.}(2004)\citenamefont
  {Cavagna}, \citenamefont {Giardina},\ and\ \citenamefont
  {Parisi}}]{cavagna:04}%
  \BibitemOpen
  \bibfield  {author} {\bibinfo {author} {\bibfnamefont {Andrea}\ \bibnamefont
  {Cavagna}}, \bibinfo {author} {\bibfnamefont {Irene}\ \bibnamefont
  {Giardina}}, \ and\ \bibinfo {author} {\bibfnamefont {Giorgio}\ \bibnamefont
  {Parisi}},\ }\bibfield  {title} {\enquote {\bibinfo {title} {{Numerical Study
  of Metastable States in Ising Spin Glasses}},}\ }\href@noop {} {\bibfield
  {journal} {\bibinfo  {journal} {Phys. Rev. Lett.}\ }\textbf {\bibinfo
  {volume} {92}},\ \bibinfo {pages} {120603} (\bibinfo {year}
  {2004})}\BibitemShut {NoStop}%
\end{thebibliography}%

\end{document}